\documentclass[10pt]{article}
\usepackage{url}
\newcommand{\noi}{\noindent}
\usepackage{perpage}
\MakePerPage{footnote}

\usepackage{graphicx}% Include figure files

\begin{document}
\bibliographystyle{osajnl}

\title{Einstein as armchair detective: The case of stimulated radiation}
\author{Vasant Natarajan\thanks{email: vasant@physics.iisc.ernet.in} \\
{\small \em Department of Physics, Indian Institute of Science, Bangalore 560 012, India}}

\maketitle

\begin{abstract}
Einstein was in many ways like a detective on a mystery trail, though in his case he was on
the trail of nature's mysteries and not some murder mystery! And like all good detectives he
had a style. It consisted of taking facts that he knew were correct and forcing nature into a
situation that would contradict this established truth. In this process she would be forced to
reveal some new truths. Einstein's 1917 paper on the quantum theory of radiation is a classic
example of this style and enabled him to predict the existence of stimulated radiation starting
from an analysis of thermodynamic equilibrium between matter and radiation.

\vspace{2mm}


\end{abstract}

Einstein is rightly regarded as one of the greatest scientific geniuses of all time. Perhaps the
most amazing and awe-inspiring feature of his work was that he was an ``armchair'' scientist,
not a scientist who spent long hours in a darkened laboratory conducting delicate
experiments, but one who performed {\it gedanken} (thought) experiments while sitting in
his favourite chair that nevertheless advanced our understanding of nature by leaps and
bounds. Two of his greatest contributions are the special theory of relativity and the general
theory of relativity, both abstract creations of his remarkable intellect. They stand out as
scientific revolutions that completely changed our perceptions of nature -- of space and time
in the case of the special theory and of gravity in the case of the general theory. It might be
argued that the special theory of relativity was necessitated by experimental facts such as the
constancy of the speed of light, but the general theory was almost completely a product of
Einstein's imagination. For a person to have achieved one revolution in his lifetime is great
enough, but two revolutions seems quite supernatural.

But is it really so magical? While it is certain that Einstein was a one-of-a-kind genius, is it at
least possible to understand the way in which his mind tackled these problems? I think the
answer is yes, because deep inside Einstein was like a detective hot on a mystery trail, of
course not one solving murder mysteries but one trying to unravel the mysteries of nature.
Any keen follower of murder mysteries knows that there are two types of detectives: those
who get down on their hands and knees looking for some microscopic piece of clinching
evidence at the scene of the crime, and the second type of ``armchair detectives'' who seem to
arrive at the solution just by thinking logically about the possibilities. Einstein was most
certainly of the second kind, and true to this breed, he had his own {\it modus operandi}. In
simple terms, his technique was to imagine nature in a situation where she contradicted
established truths, and revealed new truths in the process. As a case in point, we will look at
Einstein's 1917 paper titled ``On the quantum theory of radiation'' where he predicted the
existence of stimulated emission. While Einstein will always be remembered for his
revolutionary relativity theories, his contributions to the early quantum theory are certainly of
the highest calibre and the 1917 paper is a classic.

It is useful to first set the paper in its historical perspective. By the time Einstein wrote this
paper, he had already finished most of his work on the relativity theories. He had earlier done
his doctoral thesis on Brownian motion and was a pioneer of what is now called statistical
mechanics. He was thus a master at using thermodynamic arguments. He was one of the
earliest scientists to accept Planck's radiation law and its light quantum hypothesis. He had
already used it in 1905 for his explanation of the photoelectric effect. He was also aware of
Bohr's theory of atomic spectra and Bohr's model of the atom, which gave some explanation
for why atoms emitted radiation in discrete quanta. What he did {\bf not know} in 1917 was
any of the formalism of quantum mechanics, no Schr\"{o}dinger equation and not the de
Broglie hypothesis for wave nature of particles that we learn in high school these days.
Despite this, Einstein was successful in predicting many new things in this paper.

Let us now see what Einstein's strategy in this paper is. He is attempting to understand the
interaction between atoms and radiation from a quantum mechanical perspective. For this, he
imagines a situation where a gas of atoms is in thermal equilibrium with radiation at a
temperature $T$. The temperature $T$ determines both the Maxwell-Boltzmann velocity
distribution of atoms and the radiation density $\rho$ at different frequencies through
Planck's law. He assumes that there are two quantum states of the atom $Z_n$ and $Z_m$,
whose energies are $\varepsilon_n$ and $\varepsilon_m$ respectively, and which satisfy the
inequality $\varepsilon_m>\varepsilon_n$. The relative occupancy $W$ of these states at a
temperature $T$ depends on the Boltzmann factor as follows:
%\begin{mathletters}
\begin{eqnarray}
\label{occup1}
W_{n} &=& p_{n} \exp(-\varepsilon_{n}/kt)  \\
W_{m} &=& p_{m} \exp(-\varepsilon_{m}/kt)
\label{occup2}
\end{eqnarray}
%\end{mathletters}
where $p_n$ is a number, independent of $T$ and characteristic of the atom
and its $n$th quantum state, called the degeneracy or ``weight'' of the particular state. Similarly, $p_m$ is the weight of the $m$th state.

Einstein then makes the following basic hypotheses about the laws governing the absorption
and emission of radiation:
\begin{enumerate}
\item Atoms in the upper ($m$) state make a transition to the lower ($n$) state by
spontaneous emission. The probability $dW$ that such a transition occurs in the time $dt$ is
given by:
\begin{equation}
 dW = A_m^n dt
\end{equation}
$A_n^m$ in modern terminology is called the Einstein $A$ coefficient. Since this process is
intrinsic to the system and is not driven by the radiation field, it has no dependence on the
radiation density.
\item Atoms in the lower state make a transition to the upper state by absorbing radiation.
The probability that such a transition occurs in the time $dt$ is given by:
\begin{equation}
 dW = B_n^m \rho dt    \label{prob_a}
\end{equation}
$B_m^n$ is now called the Einstein $B$ coefficient. The absorption process is driven by the
radiation field, therefore the probability is directly proportional to the radiation density $\rho$
at frequency $\nu$.
\item The two postulates above seem quite reasonable. Now comes his new postulate, that
there is a third process of radiative transition from the upper state to the lower state, namely
stimulated emission, {\it driven by the radiation field}. By analogy with the probability for
absorption, the probability for stimulated emission is:
\begin{equation}
 dW = B_m^n \rho dt   \label{prob_e}
\end{equation}
\end{enumerate}
\noindent
Einstein calls the processes in both 2 and 3 as ``changes of state due to irradiation''. We will
see below how he is forced to include postulate 3 in order to maintain thermodynamic
equilibrium.

The main requirement of thermodynamic equilibrium is that the occupancy of atomic levels
given by Eqs.\ \ref{occup1} and \ref{occup2} should not be disturbed by the absorption and emission processes
postulated above. Therefore the number of absorption processes (type 2) per unit time from
state $n$ into state $m$ should equal the number of emission processes (type 1 and 3
combined) out of state $m$ into state $n$. This is called detailed balance. Since the number
of processes from a given state occurring in a time $dt$ is given by the occupancy of that
state times the probability of a transition, the detailed balance condition is written as:
\begin{equation}
 p_n \exp(-\varepsilon_n/kT) B_n^m \rho
= p_m \exp(-\varepsilon_m/kT) (B_m^n \rho + A_m^n)   \label{equil}
\end{equation}
Notice the importance of the third hypothesis about stimulated emission to make the equation
consistent. If one does not put that in, the equation becomes:
\begin{equation}
%\eqnum{$5'$}
 p_n \exp(-\varepsilon_n/kT) B_n^m \rho = p_m \exp(-\varepsilon_m/kT) A_m^n
\end{equation}
which clearly will not work. At high temperatures, when the Boltzmann factor makes the
occupancy of the two levels almost equal, the rate of absorption on the LHS increases with
temperature as the radiation density increases. But the rate of emission on the RHS does not
increase because spontaneous emission is independent of the radiation density.
Thermodynamic equilibrium will therefore not be maintained. This is vintage Einstein: he
imagines a situation that forces a contradiction with what he ``knows'', namely thermal
equilibrium, and uses it to obtain a new result, namely stimulated emission during radiative
transfer.

With the grace and confidence of an Olympic hurdler, Einstein now moves on to make
quantitative predictions based on the bold new hypothesis. First he uses the high temperature
limit to derive a relation between the coefficients for absorption and stimulated emission.
Under the reasonable assumption that $\rho \rightarrow \infty$ as $T \rightarrow \infty$, the
spontaneous emission term on the RHS of Eq.\ \ref{equil} can be neglected at high
temperatures. From this, it follows that:
\begin{equation}
 p_n B_n^m = p_m B_m^n
\end{equation}
By substituting this result in Eq.\ \ref{equil}, Einstein obtains a new, simple derivation of
Planck's law:
\begin{equation}
 \rho = \frac{A_m^n/B_m^n}{\exp[(\varepsilon_m-\varepsilon_n)/kT] -1}
\end{equation}
Notice that he will not get the correct form of this law if he did not have the stimulated
emission term in Eq.\ \ref{equil}. Another reason for him to be confident that his three
hypotheses about absorption and emission are correct. He then compares the above
expression for $\rho$ with Wien's displacement law:
\begin{equation}
 \rho = \nu^3 f(\nu/T)
\end{equation}
to obtain
\begin{equation}
 \frac{A_m^n}{ B_m^n} = \alpha \nu^3
\end{equation}
and
\begin{equation}
 \varepsilon_m-\varepsilon_n = h \nu
\end{equation}
with constants $\alpha$ and $h$. The second result is well known from the Bohr theory of
atomic spectra. Einstein is now completely sure that his three hypotheses about radiation
transfer are correct since he has been able to derive both Planck's law and Bohr's principle
based on these hypotheses.

Einstein does not stop here. He now considers how interaction with radiation affects the
atomic motion in order to see if he can predict new features of the momentum transferred by
radiation.
Earlier he had argued that thermal equilibrium demands that the occupancy of the states
remain undisturbed by interaction with radiation. Now he argues that the Maxwell-Boltzmann
velocity distribution of the atoms should not be disturbed by the interaction. In other words,
the momentum transfer during absorption and emission should result in the same statistical
distribution of velocities as obtained from collisions. From kinetic theory, we know that the
Maxwell velocity distribution results in an average kinetic energy along each direction given
by:
\begin{equation}
\frac{1}{2} M \left< v^2 \right> = \frac{1}{2} kT    \label{kinetic}
\end{equation}
This result should remain unchanged by the interaction with radiation.

To calculate the momentum change during radiative transfer, Einstein brings into play his
tremendous insight into Brownian motion. As is now well known from the Langevin
equation, he argues that the momentum of the atom undergoes two types of changes during a
short time interval $\tau$. One is a frictional or damping force arising from the radiation
pressure that systematically opposes the motion. The second is a fluctuating term arising from
the random nature of the absorption-emission process. It is well known from Brownian
motion theory that the atoms would come to rest from the damping force if the fluctuating
term were not present. Thus, if the initial momentum of the atom is $Mv$, then after a time
$\tau$, the momentum will have the value:
\begin{equation}
 Mv - Rv\tau + \Delta
\end{equation}
where the second term is the damping term and the last term is the fluctuating term. If the
velocity distribution of the atoms at temperature $T$ is to remain unchanged by this momentum
transfer process, the average of the above quantity (Eq.\ 12) must be equal to $Mv$,
and the mean values of the squares of these quantities must also be equal:
\begin{equation}
 \left< (Mv - Rv\tau + \Delta)^2 \right> = \left< (Mv)^2 \right>
\end{equation}
Since we are only interested in the systematic effect of $v$ on the momentum change due to
interaction with radiation, $v$ and $\Delta$ can be regarded as independent statistical
processes and the average of the cross term $ v \Delta $ can be neglected. This yields:
\begin{equation}
 \left< \Delta^2 \right> = 2RM\left< v^2 \right> \tau
\end{equation}
To maintain consistency with kinetic theory, the value of $\left< v^2 \right>$ in the above
equation must be the same as the one in Eq.\ \ref{kinetic}. Thus:
\begin{equation}
 \frac{\left< \Delta^2 \right>}{\tau} = 2RkT     \label{fluct}
\end{equation}
This is the equation that will tell Einstein if his hypotheses about momentum transfer are
correct. In other words, he assumes that the radiation density is given by Planck's law, and
calculates $R$ and $ \left< \Delta^2 \right> $ based on some hypotheses about momentum
transfer during radiative processes. If the hypotheses are valid, the above equation should be
satisfied identically in order not to contradict thermal equilibrium.

His main hypothesis about momentum transfer is that, if the photon behaves like a localized
packet of energy $E$, it must also carry directional momentum of $E/c$. Without going into
the details, I just outline the approach he uses for calculating $R$ and $ \left< \Delta^2
\right> $. For calculating $R$, he uses the following argument. In the laboratory frame in
which the atom has a velocity $v$, the radiation is isotropic. But in the rest frame of the
atom, the radiation is anisotropic because of the Doppler shift. This gives rise to a velocity
dependent radiation density and a velocity dependent probability of absorption and stimulated
emission (from Eqs.\ \ref{prob_a} and \ref{prob_e}). The average momentum transferred to
the atom is calculated from the modified rates of absorption stimulated emission, thus
yielding $R$. $R$ does not depend on the rate of spontaneous emission because spontaneous
emission occurs independently of the radiation field and is therefore isotropic in the rest
frame of the atom.
Calculating $ \left< \Delta^2 \right> $ is relatively simpler. If each absorption or emission
process gives a momentum kick of $E/c$ in a random direction, the mean square momentum
after $\ell$ kicks is simply $\ell \times (E/c)^2$. $\ell$ is equal to twice the number of
absorption processes taking place in the time $\tau$ since each absorption process is followed
by an emission process.
Using this approach, Einstein calculates $R$ and $ \left< \Delta^2 \right> $. He shows that
Eq.\ \ref{fluct} is satisfied identically when these values are substituted, which implies that
the velocity distribution from kinetic theory is not disturbed if and only if momentum
exchange with radiation occurs in units of $E/c$ in a definite direction.

He thus concludes the paper with the following observations. There must be three processes
for radiative transfer, namely absorption, spontaneous emission, and {\it stimulated
emission}. Each of these interactions is quantized and takes place by interaction with a single
radiation bundle. The radiation bundle (which we today call a photon) carries not only energy
of $h\nu$ but also momentum of $h\nu/c$ in a well defined direction. The momentum
transferred to the atom is in the direction of propagation for absorption and in the opposite
direction for emission. And finally, ever loyal to his dislike for randomness in physical laws
(``God does not play dice!''), he concludes that one weakness of the theory is that it leaves the
duration and direction of the spontaneous emission process to ``chance''. However, he is
quick to point out that the results obtained are still reliable and the randomness is only a
defect of the ``present state of the theory''.

What far reaching conclusions starting from an analysis of simple thermodynamic
equilibrium. This is a truly great paper in which we see two totally new predictions. First, he
predicts the existence of stimulated emission. And to top that, for the first time since Planck
introduced radiation quanta, he shows that each quantum carries well defined momentum. He
shows that the directional momentum is present even in the case of spontaneous emission.
Thus an atom cannot decay by emitting ``outgoing radiation in the form of spherical waves''
with no momentum recoil.

Today his conclusions about momentum transfer during absorption and emission of radiation
have been abundantly verified. Equally well verified is his prediction of stimulated emission
of radiation. Stimulated emission is the mechanism responsible for operation of the laser,
which is used in everything from home computers and CD players to long distance
communication systems. Stimulated emission, or more correctly stimulated scattering,
underlies our understanding of the phenomenon of Bose-Einstein condensation. It plays an
important role in the explanation of superconductivity and superfluidity. The two predictions,
momentum transfer from photons and stimulated emission, are particularly close to my heart
because they play a fundamental role in one of my areas of research, namely laser cooling of
atoms. In laser cooling, momentum transfer from laser photons is used to cool atoms to very
low temperatures of a few millionths of a degree above absolute zero. Perhaps fittingly, it is
the randomness or ``chance'' associated with the spontaneous emission process which he
disliked so much that is responsible for the entropy loss associated with cooling. In other
words, as the randomness from the atomic motion gets reduced by cooling, it gets added to
the randomness in the radiation field through the spontaneous emission process, thus
maintaining consistency with the second law of thermodynamics.

\section*{Conclusions}
We have seen how Einstein was able to use the principle of thermodynamic equilibrium to
imagine a situation where radiation and matter were in dynamical equilibrium and from that
predict new features of the radiative transfer process. As mentioned before, this was a
recurring theme in his work, a kind of {\it modus operandi} for the great ``detective''. In his
later writings, he said that he always sought one fundamental governing principle from which
he could derive results through these kind of arguments. He found such a principle for
thermodynamics, namely the second law of thermodynamics, which states that it is
impossible to build a perpetual motion machine. He showed that the second law was a
necessary and sufficient condition for deriving all the results of thermodynamics. His quest in
the last four decades of his life was to geometrize all forces of nature. In this quest, he felt
that he had indeed found the one principle that would allow him to do this uniquely, and this
was the {\it principle of relativity:}
\begin{itemize}
\item[] {\it the laws of physics must look the same to all observers no matter what their state
of motion.}
\end{itemize}
He had already used this principle to geometrize gravity in the general theory of relativity.
His attempts at geometrizing electromagnetic forces remained an unfulfilled dream, but that
is a story for another day.

\newpage
\section*{Appendix. Examples of {\it gedanken} experiments}
We present two examples of {\it gedanken} experiments that illustrate the Einstein technique
for arriving at new results. Both of these experiments yield results associated with the general
theory of relativity, but are so simple and elegant that they can be understood without any
knowledge of the complex mathematical apparatus needed for the general theory. The first
experiment is due to Einstein himself, while the second is due to Hermann Bondi. \\

\noi
{\bf Example 1.}
This is a thought experiment devised by Einstein to arrive at the conclusion that the general
theory of relativity is an extension of the special theory which requires curved spacetime, or
spacetime in which the rules of plane (Euclidean) geometry do not apply. The ``known'' facts
are the results of special theory of relativity applicable to inertial systems, and the
equivalence principle which states that inertial mass is exactly equal to gravitational mass.
Einstein's argument proceeds as follows.

\begin{figure}
\centering{\resizebox{0.5\columnwidth}{!}{\includegraphics{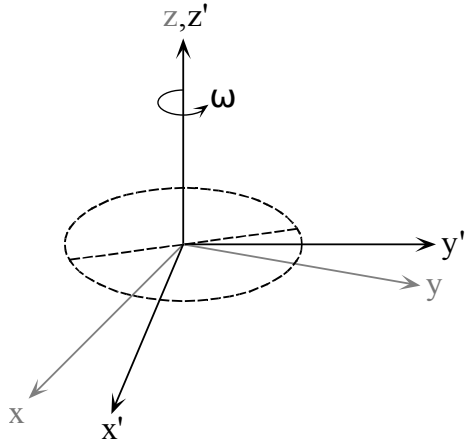}}}
\caption{Coordinate systems with relative rotation between them.}
\label{coord}
\end{figure}

Imagine two observers or coordinate systems $O$ and $O'$. Let the $z'$-axis of $O'$
coincide with the $z$-axis of $O$, and let the system $O'$ rotate about the $z$-axis of $O$
with a constant angular velocity (see {\em Figure} \ref{coord}). Thus $O$ is an inertial
system where the laws of special
relativity apply, while $O'$ is a non-inertial system. Imagine a circle drawn about the origin
in the $x'y'$ plane of $O'$ with some given diameter. Imagine, further, that we have a large
number of rigid rods, all identical to each other. We lay these rods in series along the
circumference and the diameter of the circle, at rest with respect to $O'$. If the number of
rods along the circumference is $U$ and the number of rods along the diameter is $D$, then,
if $O'$ does not rotate with respect to $O$, we have (from plane geometry)
\[ \frac{U}{D} = \pi   \]
However, if $O'$ rotates, we get a different result. We know from special relativity that,
relative to $O$, the rods on the circumference undergo Lorentz contraction while the rods
along the diameter do not undergo this contraction (the relative motion is perpendicular to the
diameter). Therefore, we are led to the unavoidable conclusion that
\[ \frac{U}{D} > \pi   \]
{\it i.e.\ }the laws of configuration of rigid bodies with respect to $O'$ is not in accordance
with plane geometry. If, further, we place two identical clocks, at rest with respect to $O'$,
one at the periphery and one at the centre of the circle, then with respect to $O$ the clock at
the periphery will go slower than the clock at the centre (from special relativity, moving
clocks go slower). A similar conclusion will be reached by $O'$, {\it i.e.\ }the two clocks go
at different rates.

We thus see that space and time cannot be defined with respect to $O'$ as they were defined
in special theory of relativity for inertial systems. But, according to the equivalence principle,
$O'$ can also be considered a system at rest with respect to which there is a gravitational
field (corresponding to the centrifugal force field and the Coriolis force field). We therefore
arrive at the following remarkable result: the gravitational field influences and even
determines the geometry of the space-time continuum, and this geometry is not Euclidean.
From this conclusion, Einstein goes on to develop a curved spacetime theory of gravitation.  \\

\noi
{\bf Example 2.}
This example illustrates the use of a thought experiment to calculate the difference in rates
between two clocks placed at different gravitational potentials, called the gravitational
redshift. We have already seen in the first example how the rate of the clock at the periphery
differs from the rate of the clock at the centre. Here, we derive a quantitative value
for this difference using an Einstein-like {\it gedanken} experiment, first conceived by Bondi.
The ``known'' things are the second law of thermodynamics and the special relativistic
energy-mass relationship, $E=mc^2$. The argument proceeds as follows.

\begin{figure}
\centering{\resizebox{0.5\columnwidth}{!}{\includegraphics{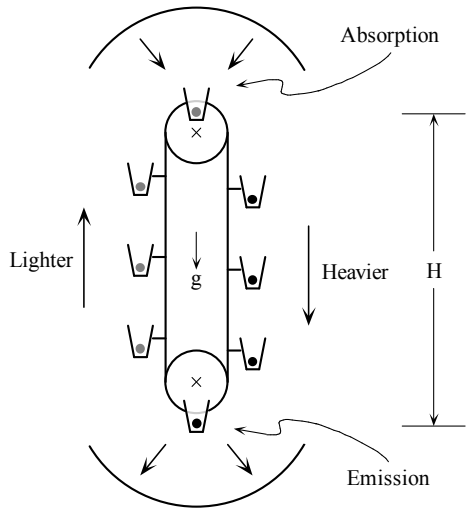}}}
\caption{
Bondi's perpetual motion machine. The buckets on the right side contain atoms that have
higher energy and are thus heavier than the atoms on the left side. When a bucket reaches the
bottom, the atom inside emits a photon which is absorbed by the corresponding atom in the
top bucket. The heavier buckets on the right keep falling down in the gravitational field and
their gravitational energy can be converted to useful work. The resolution to the paradox is
that the photon absorbed at the top has a lower frequency than the photon emitted at the
bottom.
}
\label{bondi}
\end{figure}

Imagine a series of buckets on a frictionless pulley system, as shown in {\em Figure} \ref{bondi}. Each
bucket contains an atom capable of absorbing or emitting a photon of energy $h\nu$. The
system is in a uniform gravitational field with acceleration $g$. If the photon frequency were
unaffected by the gravitational field, we can operate the system as a perpetual motion
machine in the following way. Imagine that the pulleys rotate clockwise and that all the
atoms on the left are in the ground state and the atoms on the right are in the excited state.
The lifetime in the excited state is such that, on average,
every time a bucket reaches the bottom the atom inside decays to the ground
state and emits a photon. Suitable reflectors direct this photon to the
corresponding bucket at the top so that the atom inside absorbs the photon
and goes into the excited state.
All the excited state atoms on the right have
more energy and, from the relation $E=mc^2$, are therefore heavier by an amount $\Delta
m=h\nu /c^2$. The heavier masses are accelerated down by the gravitational field
and the system remains in perpetual motion. The excess gravitational potential energy can be
converted to unlimited useful work, in violation of the second law of thermodynamics.

The solution to the paradox lies in postulating that the frequency of the photon emitted by the
atom at the bottom is not the same as the frequency of the photon when it reaches the top. Let
the two frequencies be $\nu$ and $\nu'$ respectively. Then the additional mass for the atom
at the top by absorbing a photon of frequency $\nu'$ is $h\nu' /c^2$, and the potential energy
of this excess mass at a height $H$ between the two buckets is $h\nu'/c^2 \times gH$. To
maintain consistency with the second law of thermodynamics, this excess energy should
exactly compensate for the loss in energy of the photon as its frequency changes from $\nu$
to $\nu'$:
\[ \frac{h \nu'}{c^2} gH = h (\nu - \nu') \]
which yields
\[ \frac{\nu'-\nu}{\nu'} = - \frac{g H}{c^2}    \]
{\it i.e.\ }the relative frequency shift is given by $gH/c^2$ and is negative (redshift) at the
location where the gravitational potential is higher. The shift can be understood from the fact
that the photon is also affected by the gravitational field and therefore loses energy as it
climbs up the potential. Since the photon always travels at the speed $c$, it loses energy by
changing its frequency. This result explains why, in the first example, the clock at the centre
goes slower than the clock in the periphery according to $O'$. With respect to $O'$, there is a
gravitational field (corresponding to the centrifugal force) pointing away from the centre. The
clock at the centre is at a higher gravitational potential and hence goes slower.

The gravitational redshift on the surface of the earth is very tiny at any reasonable height, but
it was experimentally verified in a remarkable experiment by Pound and Rebka in 1959. They
measured the frequency shift between the top and bottom of a building at Harvard University,
a height difference of about 25 m. The relative frequency shift measured was a tiny 3 parts in
$10^{14}$, consistent with the above calculation!

\section*{Author introduction:}
Vasant Natarajan is at the Department of Physics, Indian Institute of Science. His current research involved trapping of atoms to carry out high precision tests of fundamental physics. He has earlier worked on high precision mass spectrometry and on the focussing of atomic beams by laser fields.

\end{document}